# Non-discrimination law in Europe: a primer for non-lawyers

Frederik Zuiderveen Borgesius,[1] Nina Baranowska,[2] Philipp Hacker,[3] & Alessandro Fabris[4]

Working draft 1. We'd love to hear your comments and suggestions. Thank you!

[1] iHub, Radboud University, The Netherlands, frederikzb@cs.ru.nl
[2] iHub, Radboud University, The Netherlands, nina.baranowska@ru.nl
[3] European University Viadrina Frankfurt (Oder), Germany, hacker@europa-uni.de
[4] Max Planck Institute for Security and Privacy (MPI-SP), Germany, alessandro.fabris@mpi-sp.org

## ABSTRACT

This brief paper provides an introduction to non-discrimination law in Europe. It answers the questions: What are the key characteristics of non-discrimination law in Europe, and how do the different statutes relate to one another? Our main target group is computer scientists and users of artificial intelligence (AI) interested in an introduction to non-discrimination law in Europe. Notably, non-discrimination law in Europe differs significantly from non-discrimination law in other countries, such as the US. We aim to describe the law in such a way that non-lawyers and non-European lawyers can easily grasp its contents and challenges. The paper shows that the human right to non-discrimination, to some extent, protects individuals against private actors, such as companies. We introduce the EU-wide non-discrimination rules which are included in a number of EU directives, and also explain the difference between direct and indirect discrimination. Significantly, an organization can be fined for indirect discrimination even if the company, or its AI system, discriminated by accident. The last section broadens the horizon to include bias-relevant law and cases from the GDPR, the EU AI Act, and related statutes. Finally, we give reading tips for those inclined to learn more about non-discrimination law in Europe.

## 1 INTRODUCTION

This paper provides a first introduction to non-discrimination law in Europe. We aim to describe the law such that non-lawyers and non-Europeans may follow the arguments. Our main target group is computer scientists, but the paper could also be relevant for other readers, such as non-lawyers, or lawyers who have never looked at non-discrimination law before. The paper is structured around the following questions: What are the main characteristics of non-discrimination law in Europe, and how do the different statutes relate to one another? We focus on the region of the European Union, with its 27 Member States, not the national laws of the Member States.

For ease of reading, this paper discusses the directives, rather than national implementation laws (see [1] for an overview of national non-discrimination law in the EU).

The paper contains many footnotes, as is typical for legal writing. The footnotes primarily refer to legal provisions. To follow the main story of the paper, it is not necessary to read the footnotes. We use in-text citations when we refer to literature that we think can be useful for readers. Where possible, we refer to literature that is available for free (open access). At the bottom of the paper, we provide tips for further reading about non-discrimination law in Europe.

The paper introduces non-discrimination law and discusses overarching themes. We give rough summaries of legal concepts. If readers must work with non-discrimination law in Europe in practice, they should seek specialized legal advice.

The remainder of the paper is structured as follows. Section 2 introduces two important organizations in Europe, the Council of Europe and the European Union, and their role in non-discrimination law. Section 3 describes the non-discrimination directives of the European Union, including the scope of the directives, and direct and indirect discrimination. Section 4 explains how the GDPR, the EU AI Act, and related statutes interact with traditional non-discrimination law. Section 5 concludes.

## 2 HUMAN RIGHTS LAW IN EUROPE

In Europe, the legal regime for human rights is complicated. Two of the most important organizations for human rights in Europe are the Council of Europe and the European Union. We introduce each organization below.

### 2.1 The Council of Europe and the European Convention on Human Rights

The Council of Europe is the most important human rights organization in Europe. It is based in Strasbourg and has 46 Member States, including all 27 EU Member States. The European Convention on Human Rights is a treaty of the Council of Europe that entered into force in 1953. It was adopted in reaction to the Second World War, which showed that states can behave horribly towards citizens. All Council of Europe Member States have signed up to the European Convention on Human Rights.

The right to non-discrimination is laid down in Article 14 of the European Convention on Human Rights. Article 14 reads as follows:

'Article 14

The enjoyment of the rights and freedoms set forth in this Convention shall be secured without discrimination on any ground such as sex, race, colour, language, religion, political or other opinion, national or social origin, association with a national minority, property, birth or other status.'

The European Court of Human Rights, based in Strasburg, rules on alleged violations of the rights in the European Convention on Human Rights.[1] If somebody feels that their human rights are violated by a Member State of the Council of Europe, they should first sue their state in a national court. Once they have exhausted all possibilities in national courts, they can submit a complaint to the European Court of Human Rights. That court has developed extensive case-law on the right to non-discrimination. Member States must comply with the judgments of the European Court of Human Rights.[2]

The right to non-discrimination is also included in other human rights treaties, such as the International Convention on the Elimination of all Forms of Racial Discrimination (1965),[3] and the International Covenant on Civil and Political Rights (1966).[4] But in Europe, the European Convention on Human Rights is more relevant, mostly because of the case law of the European Court of Human Rights.

Horizontal effect of human rights

The European Court of Human Rights was originally envisioned to protect people against the state. The state has a (so called 'negative' duty not to interfere too much in people's lives. European lawyers call the relation between the state and the individual a 'vertical' relation. But the European Court of Human Rights also derives 'positive' duties for states from the European Convention on Human Rights. Hence, sometimes, the state must take action to protect people from interferences by other private actors. People cannot sue another private party under the European Convention on Human Rights. But people can complain to the European Court of Human Rights if the state does not adequately protect their rights against infringements by other non-state actors e.g., private companies. This way, the Convention's rights also have a certain 'horizontal' effect.

The practice of applying human rights in horizontal relations is still developing. The European Court of Human Rights says it 'does not consider it desirable, let alone necessary, to elaborate a general theory concerning the extent to which the Convention guarantees should be extended to relations between private individuals *inter se*.'[5] We do not give a full description of this horizontal effect. Our main point is that, in Europe, human rights can be relevant for the relation between private actors in Europe. (This contrasts with, for instance, the Bill of Rights in the United States. Those rights are only relevant for the relation between the state and the individual.)

### 2.2 The European Union and its Charter of Fundamental Rights

The European Union is a different organization than the Council of Europe. Six European countries created the European Coal and Steel Community in 1951. Six years later, the European Economic Community and European Atomic Energy Community were formed. Following their merger in 1967, the three organizations operated under the name European Communities. In 1992, the name was changed again to the European Union (EU). Currently, the EU has 27 Member States.

The EU has become increasingly important for human rights, such as privacy and non-discrimination. To illustrate: the EU adopted the influential Data Protection Directive (1995),[6] later replaced by the General Data Protection Regulation, or GDPR (2016)[7].

The Court of Justice of the European Union is one of the core EU institutions and is based in Luxembourg.[8] National judges in the EU can, and in some cases must, ask the Court of Justice of the European Union for a preliminary judgment concerning the interpretation of EU law [2].[9]

---

[1] Article 19 and 34 of the European Convention on Human Rights.
[2] Article 46 of the European Convention on Human Rights.
[3] See in particular Article 1-7 of the International Convention on the Elimination of all Forms of Racial Discrimination.
[4] Article 2 and 26 of the International Covenant on Civil and Political Rights.
[5] ECtHR, VGT Verein Gegen Tierfabriken v. Switzerland, App. No. 24699/94, 28 June 2001, para. 46.
[6] Directive 95/46/EC of the European Parliament and of the Council of 24 October 1995 on the protection of individuals with regard to the processing of personal data and on the free movement of such data, OJ L 281, 23.11.1995, p. 31–50.

[7] Regulation (EU) 2016/679 of the European Parliament and of the Council of 27 April 2016 on the protection of natural persons with regard to the processing of personal data and on the free movement of such data, and repealing Directive 95/46/EC (General Data Protection Regulation) (Text with EEA relevance), OJ L 119, 4.5.2016, p. 1–88.
[8] Article 13(1) of the Treaty on European Union (consolidated version) ([2016] OJ C 202).
[9] Article 267(2) and (3) of the Treaty on the Functioning of the European Union.

In 2000, the European Union adopted the Charter of Fundamental Rights of the European Union, a document that lists the fundamental rights and freedoms recognized by the EU. (The phrases 'human rights' and 'fundamental rights' can be regarded as synonyms for our purposes[3].) Article 20 and 21 of the EU Charter concern discrimination:

'Article 20, Equality before the law

Everyone is equal before the law.'

'Article 21, Non-discrimination

1. Any discrimination based on any ground such as sex, race, colour, ethnic or social origin, genetic features, language, religion or belief, political or any other opinion, membership of a national minority, property, birth, disability, age or sexual orientation shall be prohibited.

2. Within the scope of application of the Treaty establishing the European Community and of the Treaty on European Union, and without prejudice to the special provisions of those Treaties, any discrimination on grounds of nationality shall be prohibited.'

The Charter binds the EU institutions and the Member States when they implement EU law. In certain circumstances, the Charter may apply to the relations between private parties as well (e.g., relations between a private company and an individual).[10] In practice, national laws usually consist of more concrete anti-discrimination rules that can be applied directly to resolve a discrimination case.

Roughly summarized, Article 21(1) bans discrimination on listed grounds, such as gender and ethnicity. Article 21(2) bans discrimination on the basis of nationality.

Nationality is distinguished from the other grounds of non-discrimination because of its importance for the process of European integration and the free movement of people, including workers, within the Union. Freedom of movement for workers and any discrimination based on nationality between workers of the Member States is prohibited in one of the main legal acts forming the Union (Article 45 of the Treaty of the Functioning of the European Union). However, this rule applies only among EU citizens. For instance, the law in the Netherlands may not treat workers from Germany differently than Dutch people, because Germany is an EU Member State. The situation of non-EU workers may be governed by other regulations, such as international agreements.

The Charter of Fundamental Rights of the European Union, like other human rights treaties, has an open-ended non-discrimination clause: a clause that seems to suggest that discrimination based on any ground is prohibited. Article 21 (1) says: 'Any discrimination based *on any ground such as* sex [and other explicitly enumerated grounds] shall be prohibited' (emphasis added).[11] An open-ended clause means that courts, in deciding a particular case, can distinguish other protected grounds [4]. This approach provides greater flexibility in application of non-discrimination rules and makes the Charter more versatile [5].

Nevertheless, the Charter does not prohibit all unequal treatment on all grounds. In short, some cases of unequal treatment may be justified and do not qualify as discrimination cases. For example, a hospital can (obviously) differentiate between job candidates based on medical diplomas when hiring a new surgeon.

The EU has also adopted several legal acts called 'directives', prohibiting several types of discrimination in different contexts. We discuss these directives in the next section.

# 3 THE EU NON-DISCRIMINATION DIRECTIVES

Non-discrimination law in the EU is more complicated than data protection law. For data protection law, one can read the General Data Protection Regulation (GDPR 2016), which can be seen as one statute that applies in the whole EU.

For non-discrimination law, there is no such an overarching European regulation. Hence, national differences are much greater than in data protection law.

The national non-discrimination statutes in the 27 EU Member States are partly harmonized by EU rules. The EU can adopt 'directives'. EU Member States must adopt national law to implement (give effect to) an EU directive.[12] The Treaty on the Functioning of the European Union explains: 'A directive shall be binding, as to the result to be achieved, upon each Member State to which it is addressed, but shall leave to the national authorities the choice of form and methods.'[13] Member states can adopt stricter rules than required by the non-discrimination directives. For ease of reading, however, this paper discusses the directives rather than national implementation laws (see [1]for an overview of national non-discrimination law in the EU).

In fact, the EU has adopted many non-discrimination directives. Some of them were superseded by later directives. Currently, the four most relevant non-discrimination directives are the following:

(i) The Racial Equality Directive (2000) prohibits discrimination on the basis or racial or ethnic origin in many contexts.[14]

---

[10] CJEU (Grand Chamber), 17 April 2018, Vera Egenberger v Evangelisches Werk für Diakonie und Entwicklung e.V., Case C-414/16, ECLI:EU:C:2018:257
[11] Article 21(1) of the Charter of Fundamental Rights of the European Union, emphasis added. The European Convention on Human Rights also uses an open-ended list of protected characteristics in its non-discrimination provision: Article 14 of the European Convention on Human Rights.

[12] A 'Regulation' (such as the GDPR) does not have to be implemented in national law: 'A regulation shall have general application. It shall be binding in its entirety and directly applicable in all Member States' (Article 288 of the Treaty on the Functioning of the European Union).
[13] Article 288 of the Treaty on the Functioning of the European Union
[14] Council Directive 2000/43/EC Implementing the Principle of Equal Treatment Between Persons Irrespective of Racial or Ethnic Origin, 2000 OJ L 180/22.

(ii) The Employment Equality Directive (2000) prohibits discrimination on the grounds of religion or belief, disability, age, or sexual orientation in the employment context.[15]

(iii) The Gender Goods and Services Directive (2004) prohibits discrimination on the basis of gender in the context of the supply of goods and services.[16]

(iv) The Recast Gender Equality Directive (2006) prohibits discrimination on the basis of gender in the employment context.[17]

Together, the directives offer protection against discrimination based on six grounds, also called 'protected characteristics': age; disability; gender; religion or belief; racial or ethnic origin; and sexual orientation.

EU law requires Member States to establish equality bodies to fight discrimination based on race or ethnic origin[18] and gender[19]. In practice, many Member States set up equality bodies to tackle discrimination also on other grounds [6]. The equality bodies in Member States have different names (e.g. Defender of Rights in France, or National Office against Racial Discrimination in Italy). They may also hold mandate to tackle other human rights problems beyond discrimination cases [7]. In general, such bodies have a function that is somewhat comparable to the role of Data Protection Authorities in data protection law. They receive complaints from people who experienced discrimination and provide legal assistance; in some countries, they may also conduct independent inquiries or take offenders to court. Equality bodies also conduct research and are engaged in making policy recommendations to legislators [8].

## 3.1 Scope of the EU non-discrimination directives

The EU non-discrimination principle applies only to cases which fall under the scope of the EU law. For example, if the discriminatory treatment lacks any link to EU markets or citizens, protection against discrimination is not guaranteed by the EU law. The EU non-discrimination directives vary considerably in scope and are limited to particular sectors.

The Racial Equality Directive (2000) has the widest scope. It prohibits discrimination on the basis of racial or ethnic origin in (i) employment contexts, (ii) the welfare system, and (iii) access to good and services.

Meanwhile, the Gender Goods and Services Directive (2004) and the Recast Gender Equality Directive (2006) prohibit gender discrimination in the context of (i) employment, (ii) social security (which has a narrower scope compared to the welfare system), and (iii) access to goods and services.

The Employment Equality Directive (2000), in turn, prohibits discrimination based on sexual orientation, disability, religion or belief, and age. The Directive applies roughly summarised, to (a) conditions for access to employment, including selection criteria, recruitment conditions, and promotion; (b) access to vocational training and practical work experience; (c) employment and working conditions, including dismissals and pay; (d) membership of an organisation of workers and the benefits provided for by such organisations.

In other words, in the employment sector, directives cover all six protected characteristics, namely: racial or ethnic origin, age, disability, gender, religion or belief, and sexual orientation. On the other hand, in the context of access to goods and services, only gender and racial or ethnic origin fall under the ambit of the EU directives.

Roughly speaking, a Member State may protect people from discrimination in additional sectors not covered by the EU directives or expand the list of protected characteristics. In practice, many Member States include additional protected grounds. For example, French law introduces such grounds as 'expressing oneself in a language other than French', economic vulnerability, and gender identity. In Austria, disability of a relative, pregnancy, parenthood, and social standing also constitute protected grounds [9].

In addition to EU law, the 27 EU Member States also have their own constitutions. Most of these constitutions include a right to non-discrimination. However, constitutional principles are often phrased somewhat abstractly, and they mainly have a vertical effect (between the government and an individual). At any rate, EU law takes primacy vis-à-vis national law, the latter covering only cases to which the former does not apply.

Taking the above into account, for a technology company to comply with EU non-discrimination law, it must consider both EU law and national law in a particular country where the company will run a business. A well-meaning company could even consider all protected grounds in all sectors beyond the requirements set by the law proper.

## 3.2 Direct discrimination

To implement the right to non-discrimination, EU law distinguishes between two fundamental categories of discrimination: 'direct' and 'indirect'. Both 'direct' and 'indirect' discrimination are prohibited; however, justifications differ between direct and indirect forms of discrimination. Gladly, the four non-discrimination directives use similar definitions for those concepts. The Racial Equality Directive defines 'direct discrimination' as follows:

---

'Direct discrimination shall be taken to occur where one person is treated less favourably than another is, has been or would be treated in a comparable situation on grounds of ethnic or ethnic origin.[20],'

Hence, direct discrimination means that an organization decides against a person on the basis of a protected characteristic, such as ethnicity. For instance, if a company says that it will not recruit people of a certain ethnicity, that is an example of direct discrimination.[21] (Direct discrimination is comparable to 'disparate treatment' in United States' law.)

## 3.2 Indirect discrimination

The second category of prohibited discrimination is 'indirect discrimination'. Indirect discrimination is a more complicated concept than direct discrimination. Roughly summarized, indirect discrimination occurs when a practice is neutral at first glance but ends up discriminating against people with a certain ethnicity (or another protected characteristic). For ease of reading, we focus on ethnicity discrimination in the following. Regarding indirect discrimination, the law focuses on the effects of a practice; the intention of the alleged discriminator is not relevant. Hence, even if an organization can prove that it did not know that its AI system discriminated unfairly, that will not help the organization. It is still responsible. Indirect discrimination is comparable to 'disparate impact' in United States' law, but is not exactly the same [10-12].

For the EU, indirect discrimination is defined as follows in the Racial Equality Directive:

'Indirect discrimination shall be taken to occur where [i] an apparently neutral provision, criterion or practice [ii] would put persons of a racial or ethnic origin at a particular disadvantage compared with other persons, [iii] unless that provision, criterion or practice is objectively justified by a legitimate aim and the means of achieving that aim are appropriate and necessary.'[22]

We discuss each of the three elements of the definition of indirect discrimination in turn, and in the next section we give an example to illustrate how the definition is used in practice.

(i) What is an apparently neutral practice (or provision or criterion)? An 'apparently neutral' practice, says the Court of Justice of the European Union, means a 'practice which is worded or applied, ostensibly, in a neutral manner, that is to say, having regard to factors different from and not equivalent to the protected characteristic.'[23]

(ii) The definition of 'indirect discrimination' applies if an apparently neutral practice 'put[s] persons of a racial or ethnic origin at a particular disadvantage compared with other persons'[24]

The word 'disadvantage' must be interpreted widely, says the Court of Justice of the European Union.[25]

(iii) The third element of indirect discrimination is a lack of objective justification. The apparently neutral practice is not prohibited, if the 'practice is objectively justified by a legitimate aim and the means of achieving that aim are appropriate and necessary'.[26] Hence, this element can be broken down into two sub elements: (a) a legitimate aim, and (b) appropriate and necessary means.

To help victims of discrimination, EU law reverses the burden of proof in court cases. When somebody brings a case on discrimination to court and provides 'facts from which it may be presumed that there has been direct or indirect discrimination, it shall be for the respondent to prove that there has been no breach of the principle of equal treatment.'[27]

However, the victim must still show that discrimination is likely. Victims can use statistics to this effect.[28] However, it is often difficult for victims to obtain statistical evidence to suggest such an (indirectly) discriminatory effect (see also 4.1 and 4.2).

Below, we give a real-life example to illustrate how the ban on indirect discrimination can be applied in practice.

## 3.3 Example: Dutch life insurance case, Dazure

In 2014, the Dutch non-discrimination authority advised Dazure, an insurance company. The case does not concern an AI system. But it illustrates the steps in an analysis of alleged indirect discrimination.

The insurer offered life insurance; people could buy the life insurance online. With life insurance, the insurer pays money to the partner of the insured when the latter dies. Insurance companies want to charge higher premiums if they expect that the insured person will die early.

The insurer calculated the insurance premium based on several factors, including the consumer's postcode. Studies show that people with a higher income live longer than people with a lower income.[29] Consequently, the chance that the insurance company has to pay out earlier is greater for people with a lower income.

The insurer calculated the level of the premium based on the risk that it will have to pay out the insured amount: the lower the consumer's income, the higher the premium. However, the insurer did not ask for income directly, but used the postcode of the insured person. Having knowledge of the average income in all postcode areas in the Netherlands, the insurer wanted to offer insurance that

is easy to buy online. (The insurer also asked a few more questions, for instance, about smoking habits.)

The insurer asked the Dutch non-discrimination authority for advice on the following question: Is determining the level of the premium in this way discriminatory for people with a low-income and therefore for people with an immigrant background (non-white people)? In the Netherlands, a low income correlates with an immigrant background. We summarize how the Dutch non-discrimination authority applied the elements of the legal definition of indirect discrimination.[30]

First, the authority established that this was not a case of direct discrimination. After all, the insurer did not ask for people's ethnicity or another protected ground, but asked for people's postcode. Next, the authority assessed whether there was a case of indirect discrimination.

(i) The authority established that asking people for their postcode and using that input to calculate an insurance premium counts as an 'apparently neutral' practice.

(ii) Did that 'apparently neutral' practice 'put persons of a racial or ethnic origin at a particular disadvantage compared with other persons'?[31] Yes, concluded the authority. Postcode and low income both correlate with non-whiteness in the Netherlands. Hence, the insurer's practice led, on average, to higher prices for people with lower incomes, and thus for non-white people.

(iii) The next question is: is the apparently neutral 'practice (…) objectively justified by a legitimate aim and [are] the means of achieving that aim (…) appropriate and necessary'? The first sub-element concerns the 'legitimate aim'. The insurer said that the aim of the differentiation based on postcode was to make the insurance available at the lowest possible premium per postcode category, with as few administrative barriers as possible. The non-discrimination authority considered this aim legitimate, because it meets a real need of the insurer, and the use of the customer's postcode for differentiating the premium is not in itself discriminatory. Hence, the insurer had a legitimate aim.

The next question was: were the means of achieving that aim 'appropriate and necessary'? The non-discrimination authority did quite some research on this point. The authority asked expert advice from two academics, specializing in insurance pricing and health economics. The experts confirmed that income is a good predictor of life expectancy. The non-discrimination authority concluded that using the postcode was 'appropriate'.

Was the company's use of the postal also 'necessary'? In European law, the word 'necessary' often implies a proportionality test. The word invites the question: is the company's use of the postcode a proportional way of aiming for the company's legitimate aim?

Yes, concluded the authority. In short, the disadvantage suffered by non-white people due to the differences in premium was proportional to the objective that the insurer aims for. And there was no alternative to the use of the postcode of the consumer – or of the average income in the postcode area where the insured lives – that does not lead to indirect discrimination and is just as practical.

In sum, the authority concluded that the indirect distinction was objectively justified. Therefore, it was not prohibited to calculate the premium based on the postcode of the insured person.

The case also highlights a different issue. The EU non-discrimination directives and the Dutch non-discrimination statute only protect against discrimination on the basis of certain protected grounds, as mentioned. It was clear that people with lower income paid higher prices. But the non-discrimination authority applied existing law and concluded: 'Discrimination on the basis of income is not prohibited in the national equal treatment legislation.'[32] Some scholars, however, argue that the law should do more to protect people against such income discrimination [13].

# 4 OTHER RELEVANT LAWS

Besides non-discrimination law proper, several other EU law statutes also influence the wider context of algorithmic fairness and non-discrimination, particularly the General Data Protection Regulation (GDPR), the upcoming EU AI Act, the Consumer Rights Directive (CRD) and the Unfair Commercial Practices Directive (UCPD). This is particularly important since the enforcement of EU non-discrimination law often relies on individual litigation of injured persons and often lacks teeth – individuals tend to shy away from court trials with unclear chances of success. By contrast, the statutes considered in this section are enforced, inter alia, by public authorities, consumer agencies, or even competitors. Therefore, they have the potential to significantly raise the enforcement dynamics in the field of non-discrimination. Significantly, these provisions provide additional safeguards that complement the *outcome-based* regulatory approach of traditional non-discrimination law (a discriminatory result is illegal, irrespective of how it was obtained) with a *process-based* regime focused on the AI pipeline and internal compliance (for tensions between these regimes, see, e.g., [14]).

## 4.1 GDPR

The GDPR establishes a comprehensive framework for any kind of personal data processing concerning persons in the EU. Personal data, in this context, is understood as any information relating to an identified or identifiable natural person (Article 4(1) GDPR). Note that it is not necessary for the name or the date of birth of the person to be mentioned; any piece of information allowing the (re-)identification of an individual, even only with the help of the

---

police, suffices [15, 16] – such as IP addresses [33] or cookie identifiers [14]. The GDPR applies even if the company does not have an establishment in the EU, but offers its services to the EU customers or tracks their behaviour (for example, via cookies, see Article 3 GDPR).

Every operation processing such data – for example, for AI training or inference, or other types of data-based analyses – needs to comply with a broad set of principles and concrete rules. In the non-discrimination context, three provisions are particularly important. First, according to Article 5(1) GDPR, personal data must be accurate, up-to-date, and processed fairly. These basic principles of data processing may be violated in cases of algorithmic discrimination, for instance, if feature values in a data set are systematically skewed against a protected group, incorrect, or too old to yield adequate predictions.

Second, sensitive data, such as age, ethnic or racial origin, religion, health data, and biometric data, is protected particularly strongly under Article 9 GDPR. This has a double importance. On the one hand, processing data belonging to these categories requires explicit consent. Without such consent it is only permitted when very specific exceptions listed in Article 9(2) GDPR apply. This is supposed to bolster protection against the risk of discrimination. In its recent judgment *Meta Platforms vs. Bundeskartellamt*, the Court of Justice of the European Union (CJEU) interpreted the concept of sensitive data in a broad manner, including data points based on which Article 9 categories may merely be 'revealed', [34] i.e., confidently deduced or predicted.

On the other hand, however, it means that algorithmic fairness audits and interventions in machine learning that would specifically measure the impact of output on groups protected under Article 9 GDPR may run afoul of these very provisions [17]. Article 9 does not contain an exemption for bias correction strategies, and obtaining explicit consent from all persons in a data set will often be infeasible. This sets up a direct clash between non-discrimination and data protection law, which will now be partially resolved under the EU AI Act, as discussed below (4.2).

Third, there are specific and important additional constraints if the processing or analysis occurs in the context of what the GDPR calls 'automated decision-making'. This concerns practices in which an output is generated without meaningful involvement of a human being and that output produces legal effects or similarly affects the data subject. An example recently decided by the CJEU is credit scoring, even if the scoring agency does not take the credit decision itself, but the final decision 'draws strongly on' the score, and even if that later decision is taken by a human being.[35] Other practices are automated matching, employment, or contractual decisions. In such cases of automated decision-making, specific safeguards need to be implemented that protect, inter alia, the fundamental right to non-discrimination of data subjects (Article 22(3) GDPR). This, arguably, includes feasible state-of-the-art algorithmic fairness practices [12].

The existence of automated decision-making also gives data subjects a right to access the data and specific properties of the model, which is paramount to prove even a so-called *prima facie* case of discrimination. A *prima facie* case is the possibility of discrimination that leads to the reversal of the burden of proof, as explained under 3.2. In standard non-discrimination law, injured parties often fail to establish a statistically significant disadvantage to them for their protected groups because they lack access to the data and models. In the *Meister* case, however, the CJEU decided that potential victims of discrimination generally do not, under non-discrimination law, have a right to receive information on those aspects.[36] This paradigm has now dramatically shifted with the GDPR, as evidenced in the recent Uber and Ola cases in which drivers demanded transparency from these two ride hailing companies [18].[37] The Amsterdam District Court concluded that 'Ola must communicate the main assessment criteria and their role in the automated decision to [the drivers], so that they can understand the criteria on the basis of which the decisions were taken and they are able to check the correctness and lawfulness of the data processing'.[38] The appeals court upheld this decision[39] and ruled that AI deployers need to provide information about the main features and the weighting of those features in individual cases (local explanation), and an explanation of why these features are central [19]. This right can now be used by data subjects, and consumer or data protection organizations, to shed light on practices of algorithmic decision making, and to establish if features closely correlating with protected attributes were used in the process. Hence, it functions as a crucial first procedural step for external AI audits by potentially injured persons.

## 4.2 AI Act

The AI Act is set to come into effect in the summer of 2024, and complements the GDPR by establishing a comprehensive regulatory framework for foundation models and AI systems in high-risk applications (enumerated in Annexes II, Section A., and III AI Act). It is the EU's flagship regulation designed to address and mitigate the risks associated with the deployment of AI systems, including bias and discrimination, but fall short of tackling these issues comprehensively. Article 9 mandates that providers of AI systems establish a comprehensive risk management framework that specifically includes mechanisms for identifying and mitigating biases. Article 10 further strengthens this framework by requiring that the data sets used to train AI systems be complete, error-free, and truly representative of the target population, to the best extent possible. This article not only mandates the evaluation of these data sets for biases, but also insists on the mitigation of any detected biases with appropriate measures. The AI Act, in this way, seeks to remedy the root cause of many AI-induced biases.

---

[33] CJEU, Case C-582/14, Breyer.

[34] CJEU, Case C-252/21, Meta Platforms v Bundeskartellamt, para 73.

[35] CJEU, Case C-634/21, SCHUFA.

[36] CJEU, Case C-415/10, Meister.

[37] See, e.g., District Court of Amsterdam, Case C /13/689705/HA RK 20-258, Ola, ECLI:NL:RBAMS:2021:1019 [Ola Judgment].

[38] Ola Judgment, para. 4.52; translation according to Anton Ekker, 'Dutch court rules on data transparency for Uber and Ola drivers' (Ekker Blog) <https://ekker.legal/en/2021/03/13/dutch-Court-rules-on-data-transparency-for-uber-and-ola-drivers/> accessed on 12 December 2023.

[39] The judgement can be found here: https://uitspraken.rechtspraak.nl/#!/details?id=ECLI:NL:GHAMS:2023:804; an unofficial translation can be found here: https://5b88ae42-7f11-4060-85ff-4724bbfed648.usrfiles.com/ugd/5b88ae_de414334d89844bea61deaaebedfbbfe.pdf.

Moreover, Article 10(5) addresses the delicate balance between bias mitigation and privacy concerns mentioned in 4.1 by allowing, exceptionally, the processing of special categories of personal data to detect and correct bias in high-risk AI systems. However, it only applies to high-risk systems, and does not allow for bias correction strategies that rely on processing sensitive data in any other scenarios, which seems short-sighted.

Article 11, along with Annexes IV and IXa, requires that bias tests during the AI system's training phase and accompanying fairness metrics be documented and disclosed. Such transparency is crucial for accountability and market participants.

Finally, Article 15(3) addresses the issue of biased feedback loops, wherein biased outputs from an AI system could potentially be used as inputs for future operations, thereby creating a cycle of discrimination (e.g., Microsoft's Tay bot which descended into racist slurs). This article mandates the implementation of measures to prevent such loops: AI systems should not become echo chambers for historical and present biases.

The AI Act empowers AI Act supervising authorities as well as national non-discrimination enforcement authorities to conduct AI audits (Articles 59 and 64), which significantly enhances the regulatory oversight of AI systems to prevent discrimination.

Overall, the AI Act, therefore, not only bolsters enforcement in the non-discrimination context but, significantly, seeks to complement the outcome-based regulation of non-discrimination law – which merely prohibits discriminatory results – with a process-based regulation that intervenes at specific points in the AI pipeline to mitigate bias, such as in data curation and generation, training, and feedback loops.

## 4.3 Consumer and Unfair Competition Law

Finally, a specific variety of discrimination concerns arises from price discrimination. While we cannot provide a comprehensive overview of the law of price discrimination in this context (see, e.g., [20-23]), we would like to highlight that the modernization of EU consumer and unfair competition law introduced specific transparency provisions in this respect.

This mechanism functions in two steps. First, the EU Consumer Rights Directive now compels traders to disclose 'where applicable, that the price [of a product] was personalised on the basis of automated decision-making' (Article 6(1)(ea) CRD). [40] Hence, sellers have to attach a visible tag to any price based on personal characteristics of the consumer (Recital 46 Modernization Directive). Second, and more importantly, any omission to include such information constitutes a breach of unfair commercial practice law under Article 7(5) of the Unfair Commercial Practices Directive. Therefore, competitors, consumer and competition organizations, and (in some countries) consumer authorities may sue violators, raising enforcement pressure.

While these rules do not contain any novel material guardrails for personalized pricing, the duty to inform about it inflicts reputational cost upon entities engaging in it. Surveys show that consumers loath personalized pricing [24]. The law, in this case, cleverly uses prevailing public opinion to effectively regulate such behaviour [25]. The downside is that the law covers only automated pricing decisions. It remains difficult to elucidate if companies are effectively complying with the disclosure duty or not. Here, the novel algorithmic audits enabled by the AI Act might help to shed light on online pricing practices.

## 5 CONLUSION

This paper has outlined the fundamental aspects of non-discrimination law in Europe. It aimed to demystify its principles for computer scientists and AI users unfamiliar with legal nuances. We highlight how European non-discrimination law, distinct from its counterparts like those in the US, extends the human right to non-discrimination against actions by private entities, including corporations. By elucidating EU directives that underpin EU-wide non-discrimination rules and distinguishing between direct and indirect discrimination, we aim to provide a clear framework for understanding how these laws apply, even in cases of unintended discriminatory outcomes by AI systems.

Furthermore, we explore how non-discrimination law intersects with broader EU legislation such as the GDPR, the EU AI Act, the Consumer Rights Directive, and the Unfair Commercial Practices Directive. These statutes not only enhance the enforcement of non-discrimination laws through a variety of actors, including public authorities and consumer agencies, but also introduce a proactive approach to ensuring algorithmic fairness by focusing on the AI development process and internal compliance.

Arguably, this layered regulatory landscape offers a more robust protection against discrimination and underscores the importance of understanding these legal mechanisms for those involved in the development and deployment of AI technologies in Europe. For those seeking deeper insights, we recommend further reading on the subject below.

## 6 READING TIPS

This section provides some reading tips for readers who want to learn more.

### European law in general

J. Krommendijk & F.J. Zuiderveen Borgesius, 'How to read EU legislation?', EU Law Analysis, February 2023

---

[40] Inserted by Directive (EU) 2019/2161 of the European Parliament and of the Council of 27 November 2019 as regards the better enforcement and modernisation of Union consumer protection rules, OJ L 328/7 18.12.2019 [Modernization Directive].

http://eulawanalysis.blogspot.com/p/how-to-read-eu-legislation.html

J. Krommendijk & F.J. Zuiderveen Borgesius, 'How to read CJEU judgments: deciphering the Kirchberg oracle', EU Law Analysis, September 2022 http://eulawanalysis.blogspot.com/p/how-to-read-cjeu-judgments-deciphering.html

**Non-discrimination law**

European Union Agency for Fundamental Rights, Handbook on European non-discrimination law, 2018 edition, https://fra.europa.eu/en/publication/2018/handbook-european-non-discrimination-law-2018-edition

R. Xenidis & J. Gerards, Algorithmic discrimination in Europe: Challenges and opportunities for gender equality and non-discrimination law, 2021, https://www.equalitylaw.eu/downloads/5361-algorithmic-discrimination-in-europe-pdf-1-975

F.J. Zuiderveen Borgesius, 'Algorithmic decision-making, price discrimination, and European non-discrimination law', European Business Law Review, 31, no. 3, p. 401-422, 2020. https://works.bepress.com/frederik-zuiderveenborgesius/5/

## ACKNOWLEDGEMENTS

The authors would like to thank Marco Mauer, Sarah Großheim, Christina Christidou, and Bastian Volk for their research assistance.

This paper ("Non-discrimination law in Europe: a primer for non-lawyers") is part of the FINDHR (Fairness and Intersectional Non-Discrimination in Human Recommendation) project that received funding from the European Union's Horizon Europe research and innovation program under grant agreement No 101070212. Views and opinions expressed are however those of the author(s) only and do not necessarily reflect those of the European Union. Neither the European Union nor the granting authority can be held responsible for them.